\documentclass[doublecol]{epl2}
\usepackage{graphicx}

\title{Stochastic phenomena of synchronization in ensembles of mean-field coupled limit cycle oscillators with two native frequencies}
\shorttitle{Stochastic inter-cluster synchronization}

\author{K. Okumura\thanks{E-mail: \email{kokumura@mikan.ap.titech.ac.jp}} \and A. Ichiki \and M. Shiino}
\shortauthor{K.Okumura, A. Ichiki and M. Shiino}

\institute{                    
  Department of Physics, Faculty of Science, Tokyo Institute of Technology, 2-12-1 Oh-okayama, Meguro-ku, Tokyo 152-8551, Japan
}

\abstract{
We study effects of independent white noise on synchronization phenomena in ensembles of coupled limit cycle oscillators with different native frequencies. We consider a simple model where the ensemble consists of two inter-connected clusters with own native frequencies and mean-field couplings are introduced between intra- and inter-clusters. Taking advantage of nonlinear mean-field coupling concept together with the law of large numbers valid in the thermodynamic limit, we employ a nonlinear Fokker-Planck equation approach that turns out to be noise level-free analysis, to {\it analytically} derive the time evolution of the order parameters. Showing the occurrence of bifurcations from chaotic attractors in the deterministic limit to limit cycle ones with increasing noise intensity, we confirm the occurrence of nonequilibrium phase transitions including inter-cluster synchronization induced by external noise.
}

\begin{document}
\maketitle

\section{Introduction}
Synchronization phenomena are ubiquitous ones observed in many fields of natural sciences \cite{Pikovsky01TEXT}. In particular, synchronization phenomena of limit cycle oscillators with different native frequencies are of paramount importance in studies of nonlinear dynamic systems and neurosciences. In neurosciences, neurons in the basal ganglia, which consists of several types of neurons with different firing frequencies \cite{Bevan02}, exhibit more synchrony in Parkinson's disease than in normal state, suggesting that neural information coding associated with action selections of motor controls is closely related to the synchronization phenomena (Ref. \cite{Hammond07} and references therein). For such a reason, to study how the noise exerts its influence on the structure of synchronization will be of great importance from the viewpoint of nonlinear dynamical controls involving changes in synchrony of oscillatory systems.

Relationships between effects of noise and synchronization phenomena have been studied using various types of models. It can be intuitively supposed that the presence of noise might deteriorate the degree of synchronization of oscillatory systems. Breakdown of coherence and synchronization due to external noise in an ensemble of limit cycle oscillators has been reported \cite{Shiino85}. In contrast, the opposite phenomena of noise-induced synchronization are becoming an active field of the study of nonlinear dynamical systems \cite{Kurrer95,Kanamaru03,Wang00,Brunel99,Sakaguchi88,Bonilla92,Acebron98,Teramae04,Kawamura08}. Synchronization phenomena induced by independent noise in coupled excitable systems (including active rotator models) have been investigated both analytically \cite{Kurrer95,Kanamaru03} and numerically \cite{Wang00}. A sparsely connected network of integrate-and-fire neurons with inhibitory couplings has been found to exhibit noise induced synchronization \cite{Brunel99}. Among analytical studies on synchronization phenomena of ensembles of stochastic limit cycle oscillators is the phase reduction analysis \cite{Sakaguchi88,Bonilla92,Acebron98,Teramae04,Kawamura08}. Such type of recent studies reveal the effects of common noise on synchronization of uncoupled oscillators \cite{Teramae04} and uncoupled two populations of oscillators \cite{Kawamura08}. These studies, however, are restricted to the case with weak noise. As far as we know, there are very few papers that concern analytical studies of the relationships between noise effects and synchronization in the case of coupled general types of limit cycle oscillators.

We address the issue of synchronization phenomena of coupled limit cycle oscillators, which is supposed to be subjected to independent noise. In stochastic systems of a finite number of coupled limit cycle oscillators with external noise, each oscillator behaves randomly under the influence of noise. The corresponding Fokker-Planck equations involving all of the variables that describe the systems of many body oscillators are linear, as one of the standard master equations for Markovian dynamics. In this case, the probability densities of the systems as their solutions, in general, exhibit ergodic property to settle into equilibrium ({\it i.e.} fixed point type) probability densities for sufficiently large times. This implies that the order parameters do not oscillate.

To consider the synchronization phenomena in systems of coupled stochastic oscillators by overcoming the problem mentioned above, one may introduce the concept of taking the thermodynamic limit based on a mean-field model. Taking advantage of these issues, it is useful to employ the nonlinear Fokker-Planck equation (NFPE) approach \cite{Shiino87,Desai78,Dawson83,Frank05}. In general, probability densities of the limit cycle oscillator systems might be multimodal under the influence of weak external noise. In some cases, however, Gaussian approximation has been shown to be useful to qualitatively understand phase transitions involving bifurcations from non-oscillatory to oscillatory states with changes in noise intensity \cite{Kawai04}.

Another way to avoid the difficulty of dealing with multimodal probability densities is to restrict models to some extent. Assuming that each element in the system is essentially governed by the vector field with a quadratic potential except for nonlinear terms representing mean-field couplings with other elements, one can take advantage of using NFPEs to exactly derive the time evolution of the order parameters in the thermodynamic limit for systems of nonlinearly mean-field coupled oscillators. The NFPEs, which may exhibit bifurcations, have turned out to be useful for studying the occurrence of chaos-nonchaos phase transitions \cite{Shiino01,Ichiki07,Shiino07}.

In this letter, we study the influence of external noise on synchronization in globally nonlinear coupled oscillators with two native frequencies. We focus our attention on the inter-cluster synchronization phenomena. Using the NFPE of mean-field coupled oscillators, we derive the time evolution of the order parameters without any approximations. Investigating so obtained ordinary differential equations for the order parameters instead of a set of the Langevin equations, we confirm the occurrence of nonequilibrium phase transitions involving changes from synchronous to asynchronous states with changes in noise strength. Part of this work has been briefly reported in the conference proceedings \cite{Okumura10a}.

\section{Model and nonlinear Fokker-Planck equation approach}
\begin{figure}[tb]
  \begin{center} 
  \includegraphics[scale=0.60]{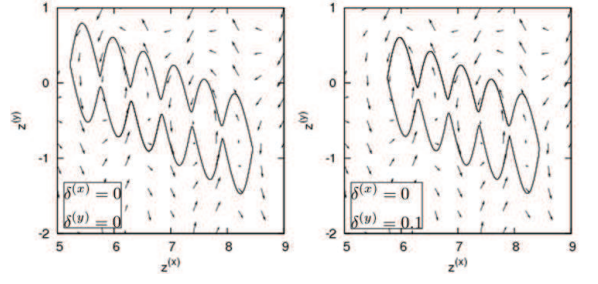}
  \end{center}
  \caption{The limit cycle attractors exhibited by the basic units. The limit cycles show ``loops'' under some parameter settings. By adjusting the detuning parameter $\delta^{(y)}$, we can change the native frequency and/or the number of loops of the limit cycle. } \label{fig:2D-ANN2_BasicUnits_deltay}
\end{figure}

To study globally coupled limit cycle oscillators with different native frequencies, first let us consider basic limit cycle oscillator units. We define a basic two dimensional oscillator of $z^{(x)}, z^{(y)}$ with the form 
$\textrm{d}z^{(x)}/\textrm{d}t = -a^{(x)}z^{(x)}+J^{(x)}F^{(x)}(b^{(x,x)}z^{(x)}+b^{(x,y)}z^{(y)})$, 
$\textrm{d}z^{(y)}/\textrm{d}t = -a^{(y)}z^{(y)}+J^{(y)}F^{(y)}(b^{(y,x)}z^{(x)}+b^{(y,y)}z^{(y)})$, 
where $a^{(\mu)}$, $b^{(\mu,\nu)}$, $J^{(\mu)}$ $(\mu=x,y)$ are constants and $F^{(\mu)}(\cdot)$ are functions specifying the nonlinearity that correspond to transfer functions in analog neural networks \cite{Marcus90,Shiino92,Shiino93,Kuhn93}. The oscillator can be of limit cycle if the functions are appropriately chosen. We here specify the nonlinear functions as $F^{(\mu)}(x)=\sin x$. Conducting a linear stability analysis of fixed points, we verify the occurrence of Hopf bifurcations under certain conditions to bring about limit cycle attractors. We set the parameter values as follows: $a^{(x)}=0.5$, $a^{(y)}=1.0$, $b^{(x,x)}=0.5$, $b^{(x,y)}=1.5$, $b^{(y,x)}=-12.0$, $b^{(y,y)}=-1.0$, $J^{(x)}=18.0$, $J^{(y)}=30.0$. To describe limit cycle oscillators with different native frequencies, we introduce detuning parameter $\delta^{(y)}$ such that $a^{(y)}=1.0+\delta^{(y)}$. The limit cycle attractors appear for $\delta^{(y)} \in (-0.4485, 19.35)$ (Fig. \ref{fig:2D-ANN2_BasicUnits_deltay}). 

We then consider a system of coupled limit cycle oscillators consisting of two clusters. The oscillators are assumed to be coupled via nonlinear global interactions and subjected to independent additive and multiplicative external noise. The model dynamics we are concerned with is described by a set of Langevin equations using the basic oscillator units mentioned above \cite{Shiino01,Ichiki07,Shiino07,Okumura10a}:\\
for site $i$ in cluster 1 $(i =1,\cdots,N_{1})$, 
\begin{eqnarray}
\frac{\textrm{d}z_{1i}^{(x)}}{\textrm{d}t} = 
-a_{1}^{(x)}z_{1i}^{(x)}+\displaystyle \sum_{j=1}^{N_1}J_{ij}^{(1,x)}F_{1}^{(x)}(b_{1}^{(x,x)}z_{1j}^{(x)}+b_{1}^{(x,y)}z_{1j}^{(y)}) \nonumber\\
+\epsilon\sum_{k=1}^{N_2}K_{ik}^{(2,x)}F_{2}^{(x)}(b_{2}^{(x,x)}z_{2k}^{(x)}+b_{2}^{(x,y)}z_{2k}^{(y)})+\eta_{1i}^{(x)}(t),\label{eq:2D-ANN2_1x}\\
\frac{\textrm{d}z_{1i}^{(y)}}{\textrm{d}t} = 
-a_{1}^{(y)}z_{1i}^{(y)}+\displaystyle \sum_{j=1}^{N_1}J_{ij}^{(1,y)}F_{1}^{(y)}(b_{1}^{(y,x)}z_{1j}^{(x)}+b_{1}^{(y,y)}z_{1j}^{(y)}) \nonumber\\
+\epsilon\sum_{k=1}^{N_2}K_{ik}^{(2,y)}F_{2}^{(y)}(b_{2}^{(y,x)}z_{2k}^{(x)}+b_{2}^{(y,y)}z_{2k}^{(y)})+\eta_{1i}^{(y)}(t), \label{eq:2D-ANN2_1y}
\end{eqnarray}
for site $i$ in cluster 2 $(i =1,\cdots,N_{2})$, 
\begin{eqnarray}
\frac{\textrm{d}z_{2i}^{(x)}}{\textrm{d}t} = 
-a_{2}^{(x)}z_{2i}^{(x)}+\displaystyle \sum_{j=1}^{N_2}J_{ij}^{(2,x)}F_{2}^{(x)}(b_{2}^{(x,x)}z_{2j}^{(x)}+b_{2}^{(x,y)}z_{2j}^{(y)}) \nonumber\\
+\epsilon\sum_{k=1}^{N_1}K_{ik}^{(1,x)}F_{1}^{(x)}(b_{1}^{(x,x)}z_{1k}^{(x)}+b_{1}^{(x,y)}z_{1k}^{(y)})+\eta_{2i}^{(x)}(t),\label{eq:2D-ANN2_2x}\\
\frac{\textrm{d}z_{2i}^{(y)}}{\textrm{d}t} = 
-a_{2}^{(y)}z_{2i}^{(y)}+\displaystyle \sum_{j=1}^{N_2}J_{ij}^{(2,y)}F_{2}^{(y)}(b_{2}^{(y,x)}z_{2j}^{(x)}+b_{2}^{(y,y)}z_{2j}^{(y)}) \nonumber\\
+\epsilon\sum_{k=1}^{N_1}K_{ik}^{(1,y)}F_{1}^{(y)}(b_{1}^{(y,x)}z_{1k}^{(x)}+b_{1}^{(y,y)}z_{1k}^{(y)})+\eta_{2i}^{(y)}(t), \label{eq:2D-ANN2_2y}
\end{eqnarray}
where $z_{\alpha i}^{(\mu)} \,(\mu=x,y) \, (\alpha=1,2)$ are the real valued-dynamical variables of the 2D-oscillators at site $i$ of cluster $\alpha$ under the natural boundary conditions, $a_{\alpha}^{(\mu)},\,b_{\alpha}^{(\mu,\nu)}$ are constants and $F_{\alpha}^{(\mu)}(\cdot)$ are nonlinear coupling functions. The mean-field intra- and inter- cluster coupling strengths $J_{ij}^{(\alpha, \mu)}(t), K_{ik}^{(\alpha, \mu)}(t)$ may include noise as 
\begin{eqnarray}
J_{ij}^{(\alpha, \mu)}(t) = \displaystyle \frac{J^{(\alpha, \mu)}}{N_{\alpha}}+\xi_{ij}^{(\alpha, \mu)}(t),\nonumber\\
K_{ik}^{(\alpha, \mu)}(t) = \displaystyle \frac{K^{(\alpha, \mu)}}{N_{\alpha}}+\zeta_{ik}^{(\alpha, \mu)}(t) \label{eq:2D-ANN2_Coupling}.
\end{eqnarray}
In cluster $1$ and $2$, we postulate the different constraint parameters as
$a_{2}^{(\mu)}=a_{1}^{(\mu)}+\delta^{(\mu)},$
with $\delta^{(\mu)}$ being essentially responsible for the difference of native frequencies between the oscillators in the two clusters. We here note that under the setting of Eqs. (\ref{eq:2D-ANN2_Coupling}), the native frequency distribution corresponds to $p(\omega) = (\delta (\omega - \omega_1(a_1)) + \delta (\omega - \omega_2(a_2)) )/2$. The inter-cluster coupling strengths are controlled by $\epsilon$. This parameter may take any real constant value, without constraining to weak connections. The external noise $\eta_{\alpha i}^{(\mu)}(t),\,\xi_{ij}^{(\alpha, \mu)}(t),\,\zeta_{ik}^{(\alpha, \mu)}(t)$ are of the white Gaussian type, 
$\langle\eta_{\alpha i}^{(\mu)}(t)\rangle=0$, 
$\langle\eta_{\alpha i}^{(\mu)}(t)\eta_{\beta j}^{(\nu)}(t^{\prime})\rangle=2D^{(\mu)}\delta_{ij}\delta_{\mu\nu}\delta_{\alpha\beta}\delta(t-t^{\prime})$, 
$\langle\xi_{ij}^{(\alpha, \mu)}(t)\rangle=0$, 
$\langle\xi_{ij}^{(\alpha,\mu)}(t)\xi_{kl}^{(\beta, \nu)}(t^{\prime})\rangle=2\tilde{D}^{(\mu)} \delta_{ik}\delta_{jl}\delta_{\mu\nu}\delta_{\alpha \beta}\delta(t-t^{\prime})/N_{\alpha}$, 
$\langle\zeta_{ik}^{(\alpha, \mu)}(t)\rangle=0$, 
$\displaystyle \langle\zeta_{ik}^{(\alpha,\mu)}(t)\zeta_{lm}^{(\beta, \nu)}(t^{\prime})\rangle=2\tilde{D}^{(\mu)} \delta_{il}\delta_{km}\delta_{\mu\nu}\delta_{\alpha \beta}\delta(t-t^{\prime})/N_{\alpha}$, 
where not common but independent noise is considered and $\eta_{\alpha i}^{(\mu)}(t),\,\xi_{ij}^{(\alpha, \mu)}(t),\,\zeta_{ik}^{(\alpha, \mu)}(t)$ are also independent of each other. We note that in the absence of noise, Eqs. (\ref{eq:2D-ANN2_1x}) - (\ref{eq:2D-ANN2_2y}) with $N_1=N_2=1$ describe the behavior of coupled limit cycle oscillators with two native frequencies, which can exhibit chaos due to the inter-cluster coupling.

In the thermodynamic limit $N_{\alpha} \rightarrow \infty$, coupling strengths of Eqs. (\ref{eq:2D-ANN2_Coupling}) ensure the convergence of each mean-field coupling term that appears in the model equations (\ref{eq:2D-ANN2_1x}) - (\ref{eq:2D-ANN2_2y}). This is attributed to the law of large numbers, giving rise to the validity of the self-averaging property. Hence the NFPE \cite{Frank05} satisfied by the empirical probability density $P(t,\textrm{\boldmath $z$})$ $(\textrm{\boldmath $z$}^{T}=(z_{1}^{(x)},\,z_{1}^{(y)},\,z_{2}^{(x)},\,z_{2}^{(y)}))$ of the Langevin equations (\ref{eq:2D-ANN2_1x}) - (\ref{eq:2D-ANN2_2y}) can easily be written down in the limit $N_{\alpha} \rightarrow \infty$ \cite{Okumura10a}. We note that the empirical probability density can also be obtained for each cluster in our model setting by taking marginal distribution: 
$P_{\alpha}(t,\textrm{\boldmath $z_{\alpha}$}) = \int \textrm{d}\textrm{\boldmath $z_{\bar{\alpha}}$} P(t,\textrm{\boldmath $z$})$, where $(\textrm{\boldmath $z$}_{\alpha}^{T}=(z_{\alpha}^{(x)},\,z_{\alpha}^{(y)}))$ and $\bar{\alpha}$ denotes the counterpart of cluster $\alpha$. 
Then, the mean-field coupling terms in Eqs. (\ref{eq:2D-ANN2_1x}) - (\ref{eq:2D-ANN2_2y}) are expressed in terms of
\begin{eqnarray}
\langle F_{\alpha}^{(\mu)}\rangle\equiv\int \textrm{d}\textrm{\boldmath $z$}_{\alpha}F_{\alpha}^{(\mu)}(b_{\alpha}^{(\mu,x)}z_{\alpha}^{(x)}+b_{\alpha}^{(\mu,y)}z_{\alpha}^{(y)}) P_{\alpha}(t,\textrm{\boldmath $z$}_{\alpha}), \label{eq:2D-ANN2_LLN_F}\\
\langle F_{\alpha}^{(\mu)^{2}}\rangle\equiv\int \textrm{d}\textrm{\boldmath $z$}_{\alpha}F_{\alpha}^{(\mu)^{2}}(b_{\alpha}^{(\mu,x)}z_{\alpha}^{(x)}+b_{\alpha}^{(\mu,y)}z_{\alpha}^{(y)}) P_{\alpha}(t,\textrm{\boldmath $z$}_{\alpha}). \label{eq:2D-ANN2_LLN_FF}
\end{eqnarray}
The total number of dynamic variables of the system is consequently reduced from $2(N_1+N_2)$ to $4$ in the thermodynamic limit. Thus one has a set of the Langevin equations as 
$\textrm{d}z_{\alpha}^{(\mu)}/\textrm{d}t=-a_{\alpha}^{(\mu)}z_{\alpha}^{(\mu)}+J^{(\alpha, \mu)}\langle F_{\alpha}^{(\mu)}\rangle+\epsilon K^{(\bar{\alpha}, \mu)}\langle F_{\bar{\alpha}}^{(\mu)}\rangle+{\eta^{\prime}}_{\alpha}^{(\mu)}(t)$,
where the original noise is transformed to effective white Gaussian noise ${\eta^{\prime}}_{\alpha}^{(\mu)}$ as 
$\langle{\eta^{\prime}}_{\alpha}^{(\mu)}(t)\rangle=0$, 
$\langle{\eta^{\prime}}_{\alpha}^{(\mu)}(t){\eta^{\prime}}_{\beta}^{(\nu)}(t^{\prime})\rangle=2D_{{\tiny \textrm{eff}}\alpha}^{(\mu)}\delta_{\mu\nu}\delta_{\alpha\beta}\delta(t-t^{\prime})$, 
$D_{{\tiny \textrm{eff}}\alpha}^{(\mu)}=D^{(\mu)}+\tilde{D}^{(\mu)} ( \langle F_{\alpha}^{(\mu)^{2}}\rangle+\epsilon^{2}\langle F_{\bar{\alpha}}^{(\mu)^{2}}\rangle )$. 
The coupled NFPEs for the empirical probability densities $P_{\alpha}(t,\textrm{\boldmath $z_{\alpha}$})$ corresponding to above the set of the effective Langevin equations read 
\begin{eqnarray}
\lefteqn{ \frac{\partial}{\partial t}P_{\alpha}(t,\textrm{\boldmath $z$}_{\alpha})=- \sum_{\mu=x,y}\frac{\partial}{\partial z_{\alpha}^{(\mu)}} \bigg(-a_{\alpha}^{(\mu)}z_{\alpha}^{(\mu)}}\nonumber\\
&\, \, +J^{(\alpha, \mu)}\displaystyle \langle F_{\alpha}^{(\mu)}\rangle+\epsilon K^{(\bar{\alpha}, \mu)}\langle F_{\bar{\alpha}}^{(\mu)}\rangle-D_{{\tiny \textrm{eff}}\alpha}^{(\mu)}\frac{\partial}{\partial z_{\alpha}^{(\mu)}}\bigg)P_{\alpha}. \label{eq:2D-ANN2_NFPE}
\end{eqnarray}

A Gaussian probability density satisfies Eqs. (\ref{eq:2D-ANN2_NFPE}) as a special solution. Since the H theorem \cite{Shiino01} ensures that the probability density satisfying Eqs. (\ref{eq:2D-ANN2_NFPE}) converges to the Gaussian-form for sufficiently large times, we represent the Gaussian probability densities as
$P_{\tiny \textrm{G} \alpha}(t,\textrm{\boldmath $z$}_{\alpha}) = \exp\left[-\frac{1}{2}\textrm{\boldmath $s$}_{\alpha}^{T}C_{\alpha}^{-1}(t)\textrm{\boldmath $s$}_{\alpha}\right]/(2\pi\sqrt{\det C_{\alpha}(t)})$, 
where 
$\textrm{\boldmath $s$}_{\alpha}^{T}=(z_{\alpha}^{(x)}-\langle z_{\alpha}^{(x)}\rangle_{\tiny \textrm{G}},\,z_{\alpha}^{(y)}-\langle z_{\alpha}^{(y)}\rangle_{\tiny \textrm{G}}) \equiv (u_{\alpha}^{(x)},\,u_{\alpha}^{(y)})$, 
$C_{{\alpha} ij}(t)=\langle s_{\alpha i}s_{\alpha j}\rangle_{\tiny \textrm{G}}$ and $\langle \cdot \rangle_{\tiny \textrm{G}}$ denotes expectation over $P_{\tiny \textrm{G}}$. Now that Eqs. (\ref{eq:2D-ANN2_LLN_F}) and (\ref{eq:2D-ANN2_LLN_FF}) are described in terms of the first and second moments, one obtains a set of closed ordinary differential equations involving at most second moments. While Gaussian approximations ignore cumulants higher than the second \cite{Kawai04}, our expressions based on the Gaussian probability densities are exact in the sense that a random variable representing the state of an individual oscillator of the nonlinearly coupled system undergoes a quadratic potential, thus constituting the Ornstein-Uhlenbeck process. The time evolution of each moment is calculated from Eqs. (\ref{eq:2D-ANN2_NFPE}) and we have
\begin{eqnarray}
&&\frac{\textrm{d}\langle z_{\alpha}^{(\mu)}\rangle_{\tiny \textrm{G}}}{\textrm{d}t}=-a_{\alpha}^{(\mu)}\langle z_{\alpha}^{(\mu)}\rangle_{\tiny \textrm{G}} \nonumber\\
&& \qquad\qquad\quad +J^{(\alpha, \mu)}\langle F_{\alpha}^{(\mu)}\rangle_{\tiny \textrm{G}}+\epsilon K^{(\bar{\alpha}, \mu)}\langle F_{\bar{\alpha}}^{(\mu)}\rangle_{\tiny \textrm{G}} \label{eq:2D-ANN2_MomentEq_mean} \\
&&\frac{\textrm{d}\langle u_{\alpha}^{(\mu)^{2}}\rangle_{\tiny \textrm{G}}}{\textrm{d}t}=-2a_{\alpha}^{(\mu)}\langle u_{\alpha}^{(\mu)^{2}}\rangle_{\tiny \textrm{G}}+2D_{{\tiny \textrm{eff}}\alpha}^{(\mu)} \label{eq:2D-ANN2_MomentEq_var} \\
&&\frac{\textrm{d}\langle u_{\alpha}^{(\mu)}u_{\beta}^{(\nu)}\rangle_{\tiny \textrm{G}}}{\textrm{d}t}=-(a_{\alpha}^{(\mu)}+a_{\beta}^{(\nu)})\langle u_{\alpha}^{(\mu)}u_{\beta}^{(\nu)}\rangle_{\tiny \textrm{G}}, \label{eq:2D-ANN2_MomentEq_cov} 
\end{eqnarray}
where
$(\alpha,\beta,\mu,\nu)=(1,1,x,y)$, $(2,2,x,y)$, $(1,2,x,x)$, $(1,2,y,y)$, $(1,2,x,y)$, $(1,2,y,x)$.
Note that $\langle u_{\alpha}^{(\mu)}u_{\beta}^{(\nu)}\rangle_{\tiny \textrm{G}}\rightarrow 0 \, (t\rightarrow\infty)$, implying that the covariant components of $C_{\alpha}$ take zero in the stationary states.

For observing qualitative as well as quantitative dynamical behaviors of the system, we proceed to solve numerically the above set of equations. When the coupling functions are specified as $F_{\alpha}^{(\mu)}(x)=\sin x$, then $\langle F_{\alpha}^{(\mu)}\rangle_{\tiny \textrm{G}}$ and $\langle F_{\alpha}^{(\mu)^2}\rangle_{\tiny \textrm{G}}$ are calculated as
\begin{eqnarray}
\lefteqn{ \langle F_{\alpha}^{(\mu)}\rangle_{\tiny \textrm{G}} = \sin\left(b_{\alpha}^{(\mu,x)}\langle z_{\alpha}^{(x)}\rangle_{\tiny \textrm{G}}+b_{\alpha}^{(\mu,y)}\langle z_{\alpha}^{(y)}\rangle_{\tiny \textrm{G}}\right)} \nonumber\\
&& \times\exp\left(-\frac{b_{\alpha}^{(\mu,x)^{2}}}{2}\langle u_{\alpha}^{(x)^{2}}\rangle_{\tiny \textrm{G}}-\frac{b_{\alpha}^{(\mu,y)^{2}}}{2}\langle u_{\alpha}^{(y)^{2}}\rangle_{\tiny \textrm{G}}\right) \label{eq:2D-ANN2_Fsin},\\
\lefteqn{ \langle F_{\alpha}^{(\mu)^{2}}\rangle_{\tiny \textrm{G}} = \displaystyle \frac{1}{2}-\frac{1}{2}\cos\left(2b_{\alpha}^{(\mu,x)}\langle z_{\alpha}^{(x)}\rangle_{\tiny \textrm{G}}+2b_{\alpha}^{(\mu,y)}\langle z_{\alpha}^{(y)}\rangle_{\tiny \textrm{G}}\right)}\nonumber\\
&& \times\exp\left(-2b_{\alpha}^{(\mu,x)^{2}}\langle u_{\alpha}^{(x)^{2}}\rangle_{\tiny \textrm{G}}-2b_{\alpha}^{(\mu,y)}\langle u_{\alpha}^{(y)^{2}}\rangle_{\tiny \textrm{G}}\right) \label{eq:2D-ANN2_FFsin}.
\end{eqnarray}

\section{Nonequilibrium phase transitions and stochastic inter-cluster synchronization}
Now that the NFPE as the time evolution of our system has been reduced to a set of moment equations (\ref{eq:2D-ANN2_MomentEq_mean})-(\ref{eq:2D-ANN2_MomentEq_cov}) that constitute the order parameter equations, it will suffice to deal with them to investigate the behavior of the system. The order parameter equations (\ref{eq:2D-ANN2_MomentEq_mean})-(\ref{eq:2D-ANN2_MomentEq_cov}) together with Eqs. (\ref{eq:2D-ANN2_Fsin}) and (\ref{eq:2D-ANN2_FFsin}) are nonlinear, yielding various types of bifurcations with changes in the parameters.

For simplicity, we only treat the Langevin noise case, {\it i.e.}, $\tilde{D}^x=\tilde{D}^y=0$, and do not study inter-cluster chaotic synchronization in this paper \cite{Pecora90}. From Eqs. (\ref{eq:2D-ANN2_MomentEq_var}), the variances turn out to be constant for sufficiently large times. Since we are concerned with investigating nonequilibrium stationary states, particularly the appearance and disappearance of the inter-cluster synchronization phenomena in the thermodynamic limit, we confine ourselves to the appearance of limit cycle attractors with changes in the noise intensity. To identify attractors of limit cycle type, we employ two criteria: (a) the numerically estimated largest Lyapunov exponent (LLE) of the attractor nearly equals zero and (b) all of the relevant order parameters exhibit periodic motions with a common time period. Note that both of them exclude the case where the attractor is of torus type.

Numerical calculations were performed with the fourth-order Runge-Kutta method under the setting of parameter values: $a_{1}^{(x)}=0.5$, $a_{1}^{(y)}=1.0$, $b_{\alpha}^{(x,x)}=0.5$, $b_{\alpha}^{(x,y)}=1.5$, $b_{\alpha}^{(y,x)}=-12.0$, $b_{\alpha}^{(y,y)}=-1.0$, $J^{(\alpha, x)}=K^{(\alpha, x)}=18.0$, $J^{(\alpha, y)}=K^{(\alpha, y)}=30.0$, $a_{2}^{(x)}=0.5+\delta^{(x)}$, $a_{2}^{(y)}=1.0+\delta^{(y)}$. Nonequilibrium phase transitions involving the inter-cluster synchronization of the limit cycle oscillators are systematically investigated with changes in the inter-cluster coupling strength $\epsilon$ and the detuning parameters $\delta^{(\mu)}$ , and the Langevin noise intensity $D^{(\mu)}$. For simplicity, we take $\delta^{(x)}=D^{(y)}=0$ in what follows.

\begin{figure}[tb]
  \begin{center} 
  \includegraphics[scale=0.50,angle=-90,clip]{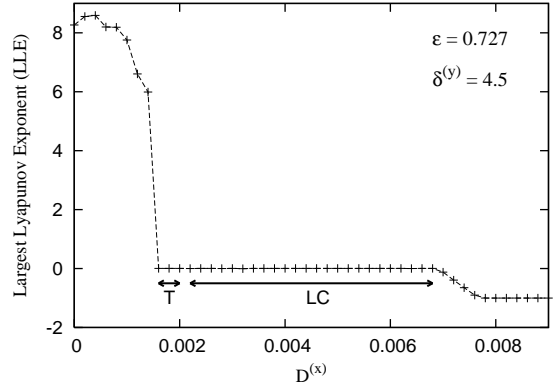}
  \end{center}
  \caption{Dependence of the largest Lyapunov exponents (LLE) on the Langevin noise intensity $D^{(x)}$, when $D^{(y)}=\delta^{(x)}=0$. Since the basic oscillators show single loop limit cycles for $\delta^{(y)}=4.5$, the system consists of coupled 6 and 1 loop limit cycles. In the deterministic limit, the LLE has a positive value, implying a chaotic attractor. For the noise intensity $D^{(x)}$ larger than a certain value, the LLE almost take zero, and the attractor changes into that of limit cycle type via torus type (denoted by LC and T, respectively). }\label{fig:2D-ANN2_LLE_deltay_Dx}
\end{figure}

We begin with investigating behaviors of the system that exhibits a chaotic attractor in the deterministic limit (Fig. \ref{fig:2D-ANN2_LLE_deltay_Dx}). The chaos originates from one pair system of different native frequencies. We note that ``deterministic limit ($D^{(x)}\rightarrow 0$)'' and ``essentially deterministic case ($D^{(x)}=0$)'' may be different dynamics. Our model is proposed so that once external noise is introduced to the system, the model constrains the form of the probability densities to the Gaussian ones for sufficiently large times due to mean-field coupling (self-organization). This indicates that all oscillators in each cluster show the same behavior in the deterministic limit. Thus chaotic attractors in the deterministic limit imply intra-cluster  synchronized chaotic behavior.

The property of worth noting of inter-cluster synchronization originates from effects of noise (Fig. \ref{fig:2D-ANN2_LLE_deltay_Dx}). With increase of the Langevin noise intensity $D^{(x)}$, the chaotic attractors with positive LLEs in weak noise limit change into those of limit cycle type via torus type, suggesting the appearance of the inter-cluster synchronization. Further increase of the noise intensity leads to negative values of the LLEs, implying fixed point type attractors. These results are qualitatively consistent with those reported by Hakim and Rappel (Figs. 1 and 2 in Ref. \cite{Hakim94}), where a set of Langevin equations were solved numerically.

\begin{figure}[tb]
  \begin{center} 
    \begin{tabular}{l}
      \resizebox{41mm}{!}{\includegraphics[scale=1.25,angle=-90,clip]{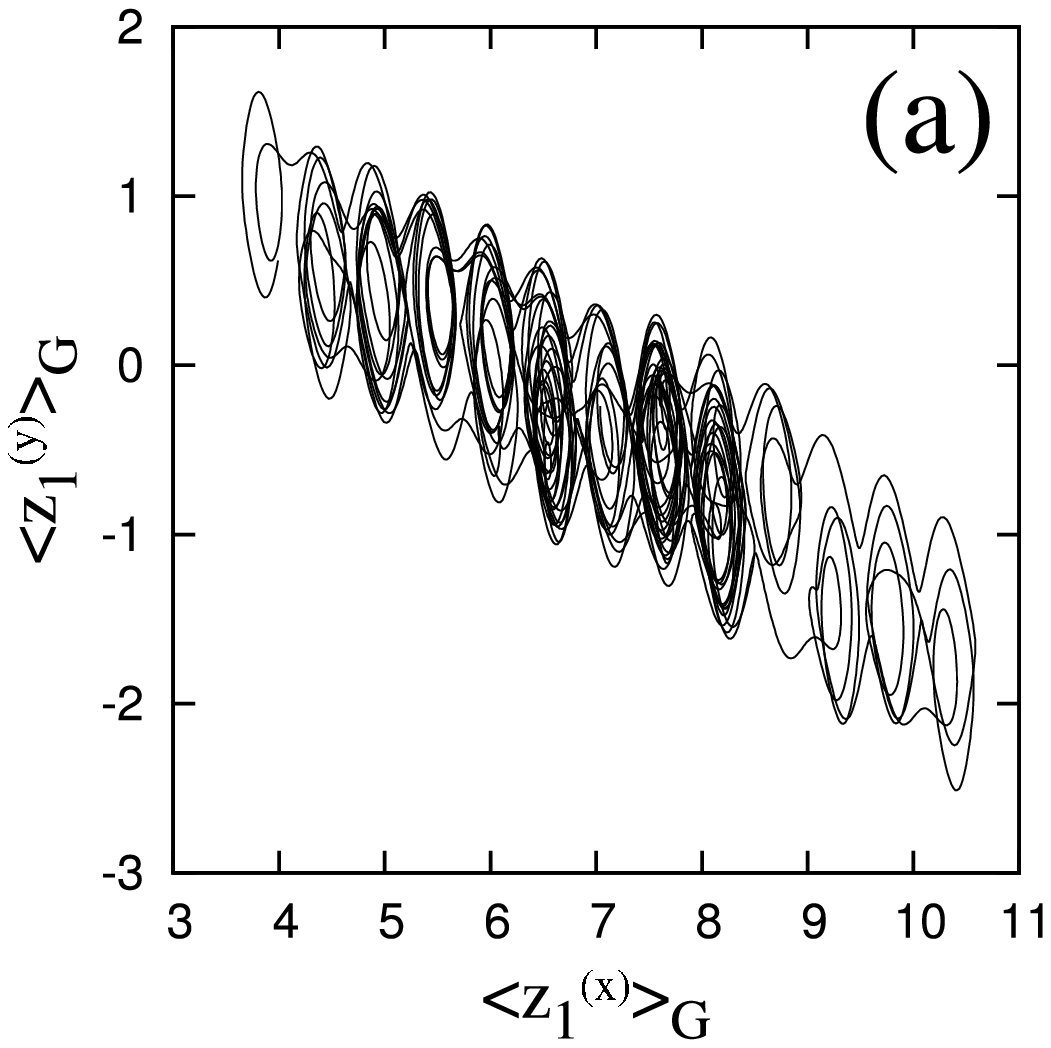}} 
      \resizebox{41mm}{!}{\includegraphics[scale=1.25,angle=-90,clip]{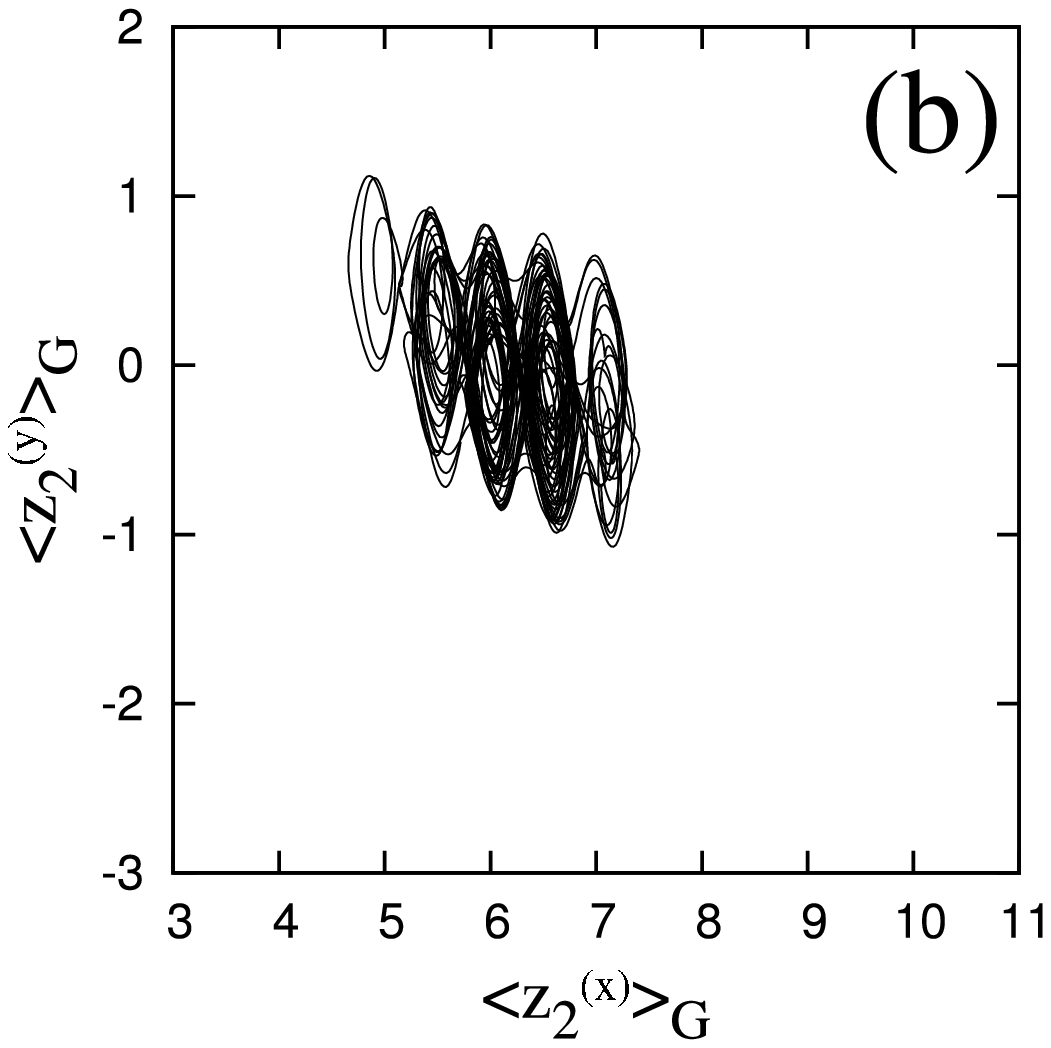}} \\
      \resizebox{41mm}{!}{\includegraphics[scale=0.30,angle=-90,clip]{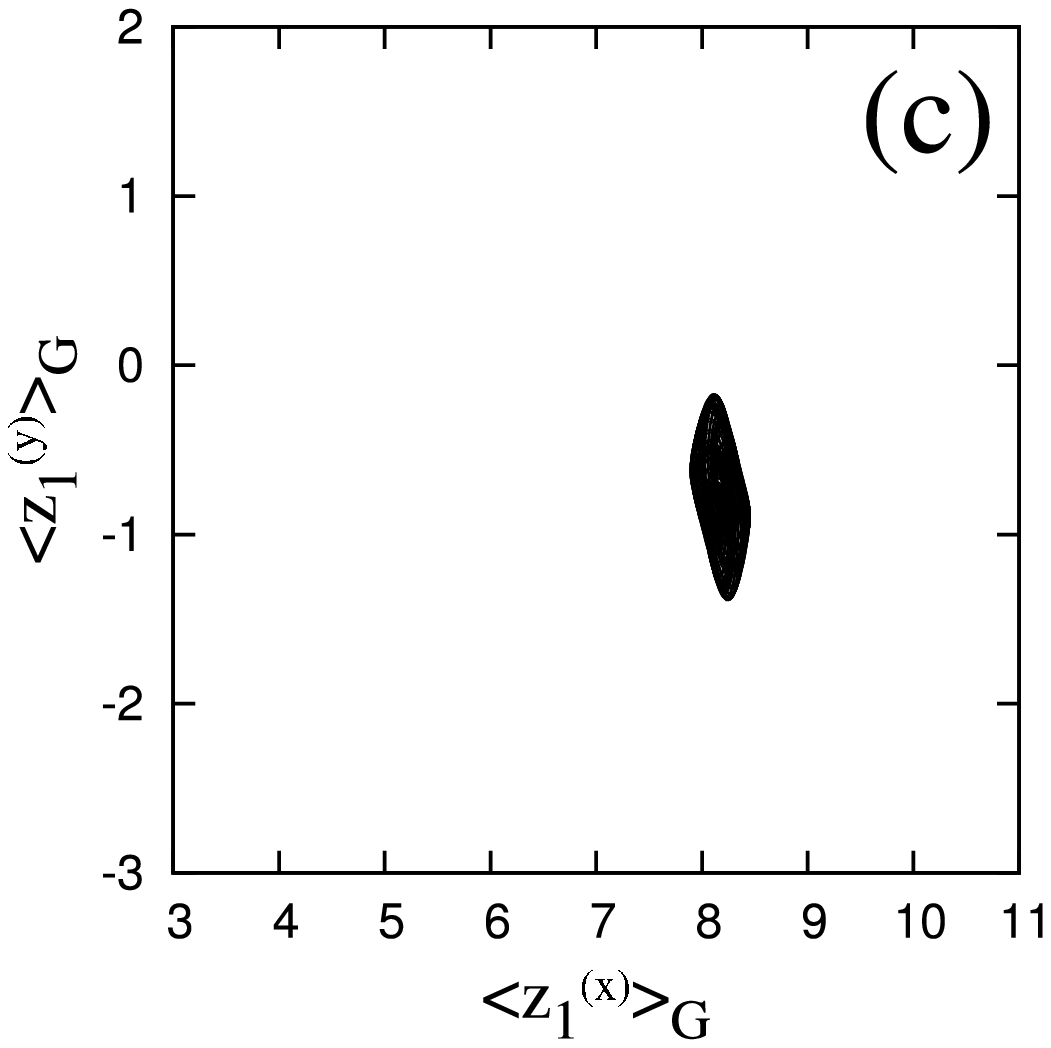}} 
      \resizebox{41mm}{!}{\includegraphics[scale=0.30,angle=-90,clip]{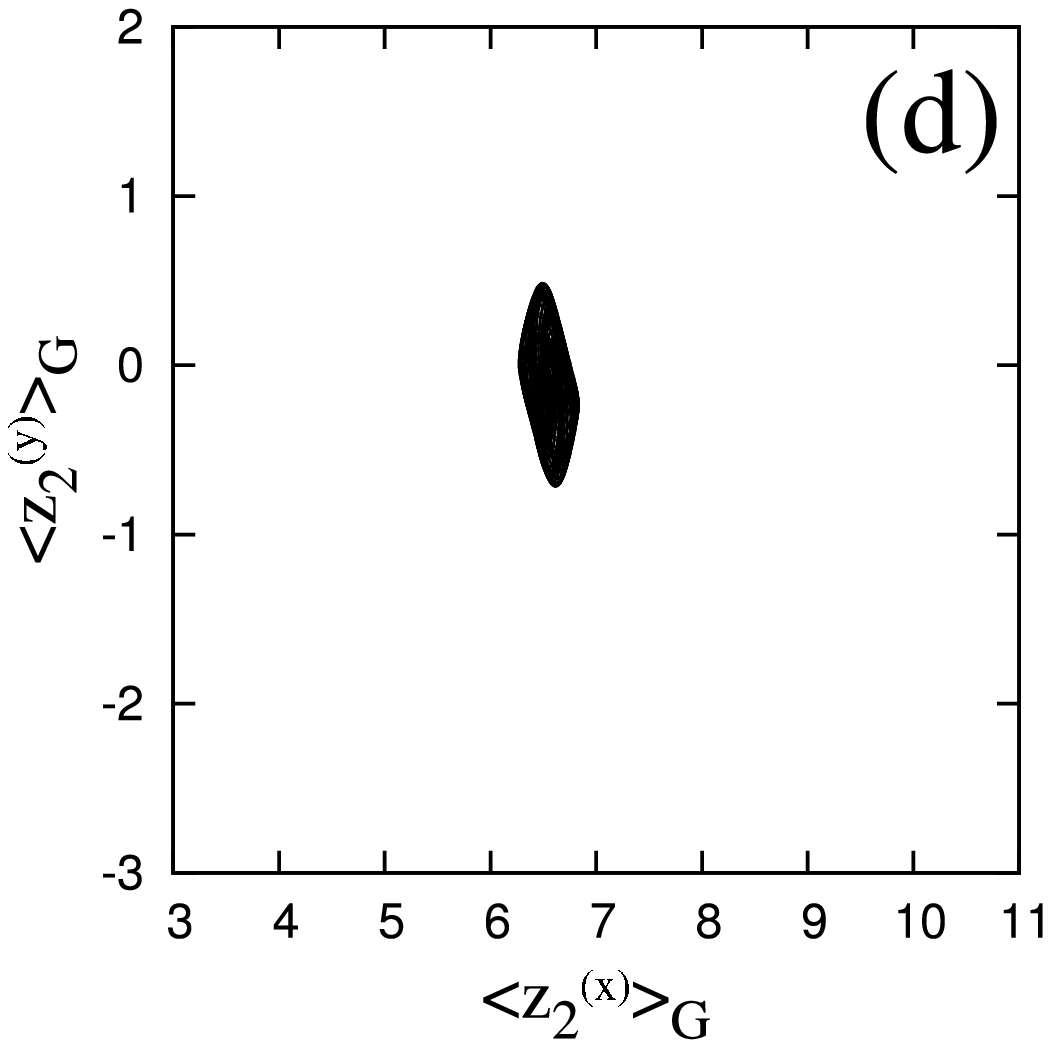}} \\
      \resizebox{41mm}{!}{\includegraphics[scale=0.20,angle=-90,clip]{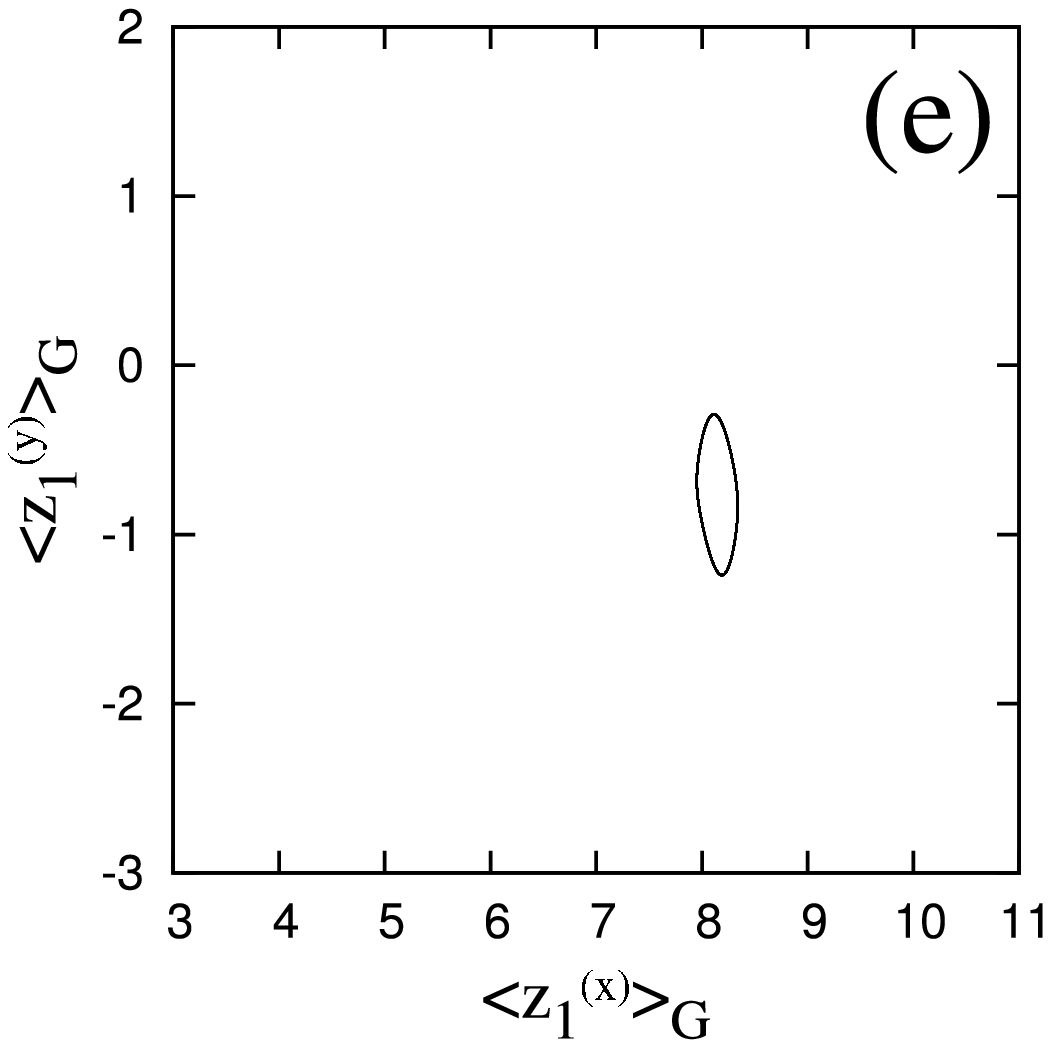}} 
      \resizebox{41mm}{!}{\includegraphics[scale=0.20,angle=-90,clip]{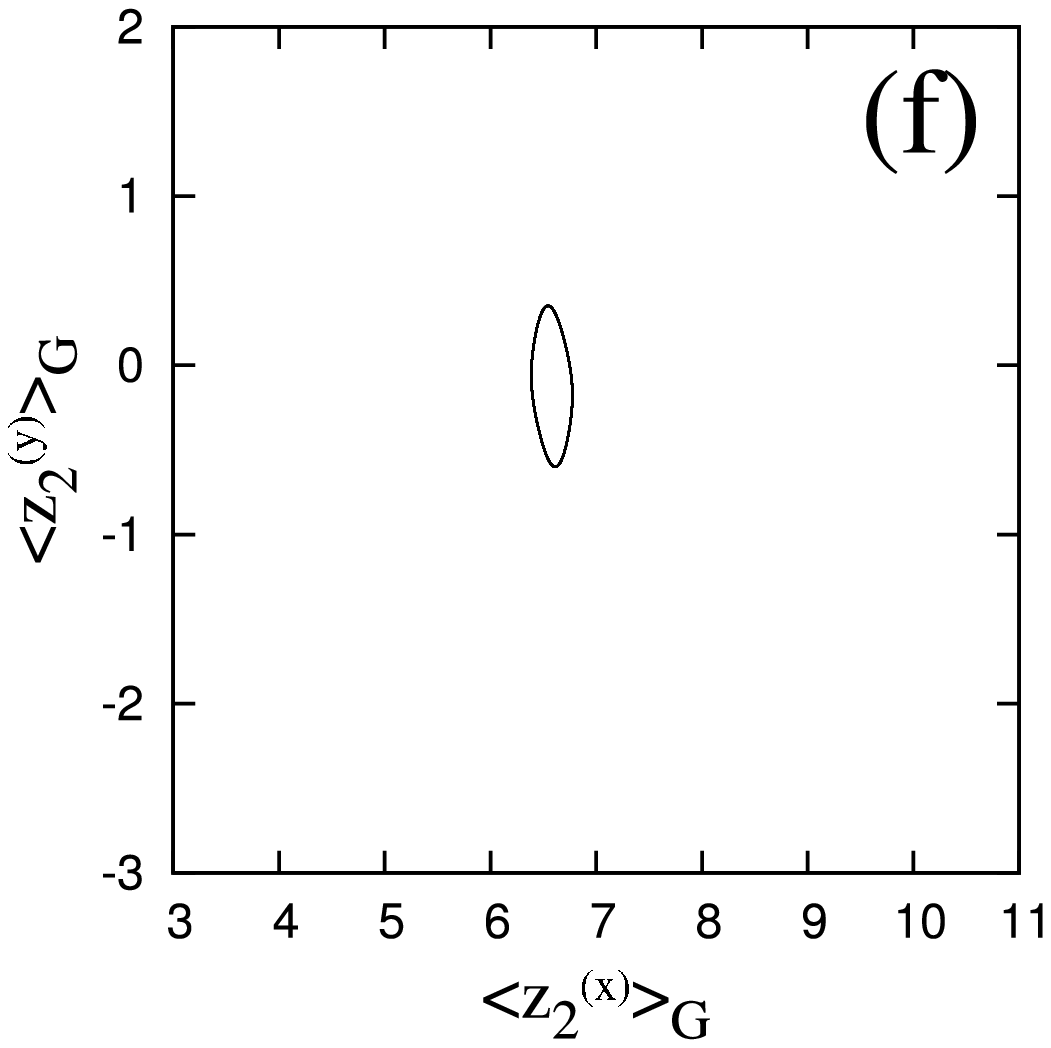}} \\
    \end{tabular}
  \end{center}
 \caption{The behavior of the order parameters of the system in phase space. The left ((a), (c), (e)) and right ((b), (d), (f)) columns show the trajectories projected to the $\langle z_{1}^{(x)}\rangle_{\tiny \textrm{G}}$-$\langle z_{1}^{(y)}\rangle_{\tiny \textrm{G}}$ and $\langle z_{2}^{(x)}\rangle_{\tiny \textrm{G}}$-$\langle z_{2}^{(y)}\rangle_{\tiny \textrm{G}}$ planes, respectively. (a), (b) In the deterministic limit $D^{(x)}=0$, it exhibits chaotic behavior. (c), (d) Certain levels of noise intensity ({\it e.g.} $D^{(x)}=0.0018$) give rise to torus type attractors. (e), (f) For larger noise intensities ({\it e.g.} $D^{(x)}=0.0022$), limit cycle attractors appear, confirming the occurrence of stochastic inter-cluster synchronization in macroscopic physical quantities. } \label{fig:2D-ANN2_PS}
\end{figure}

\begin{figure}[tb]
  \begin{center} 
    \begin{tabular}{l}
      \resizebox{41mm}{!}{\includegraphics[scale=1.25,angle=-90,clip]{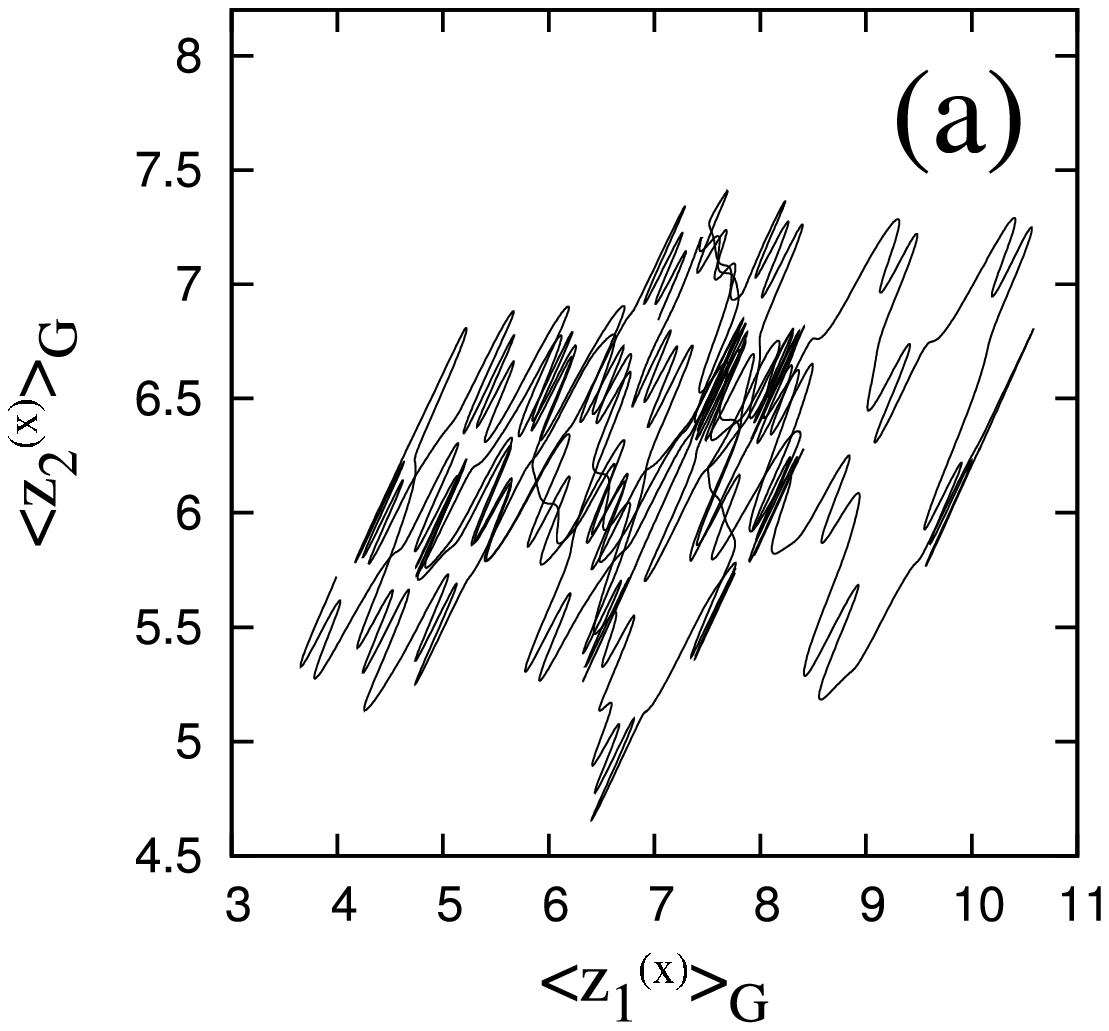}} 
      \resizebox{41mm}{!}{\includegraphics[scale=1.25,angle=-90,clip]{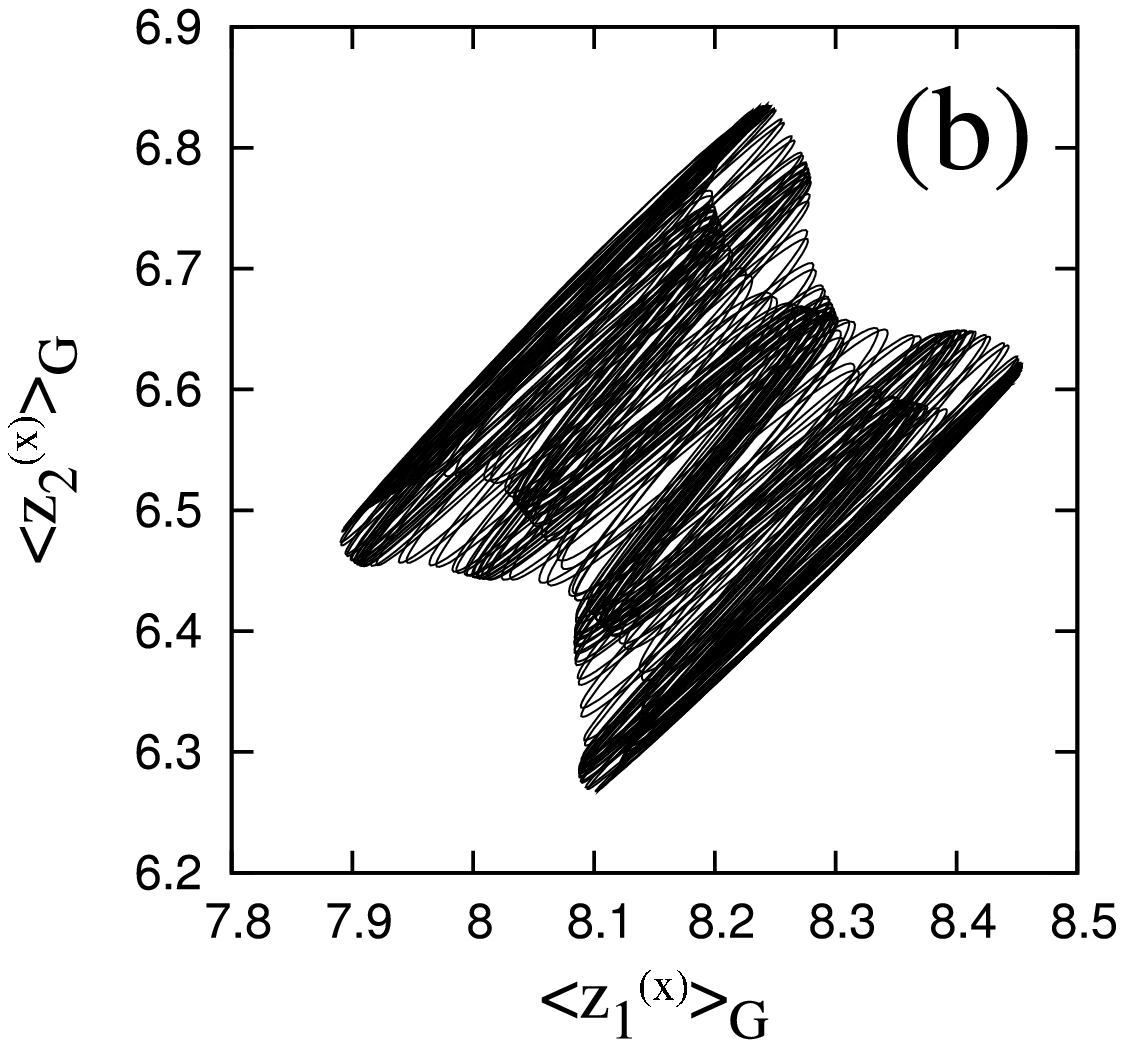}} \\
      \resizebox{41mm}{!}{\includegraphics[scale=0.30,angle=-90,clip]{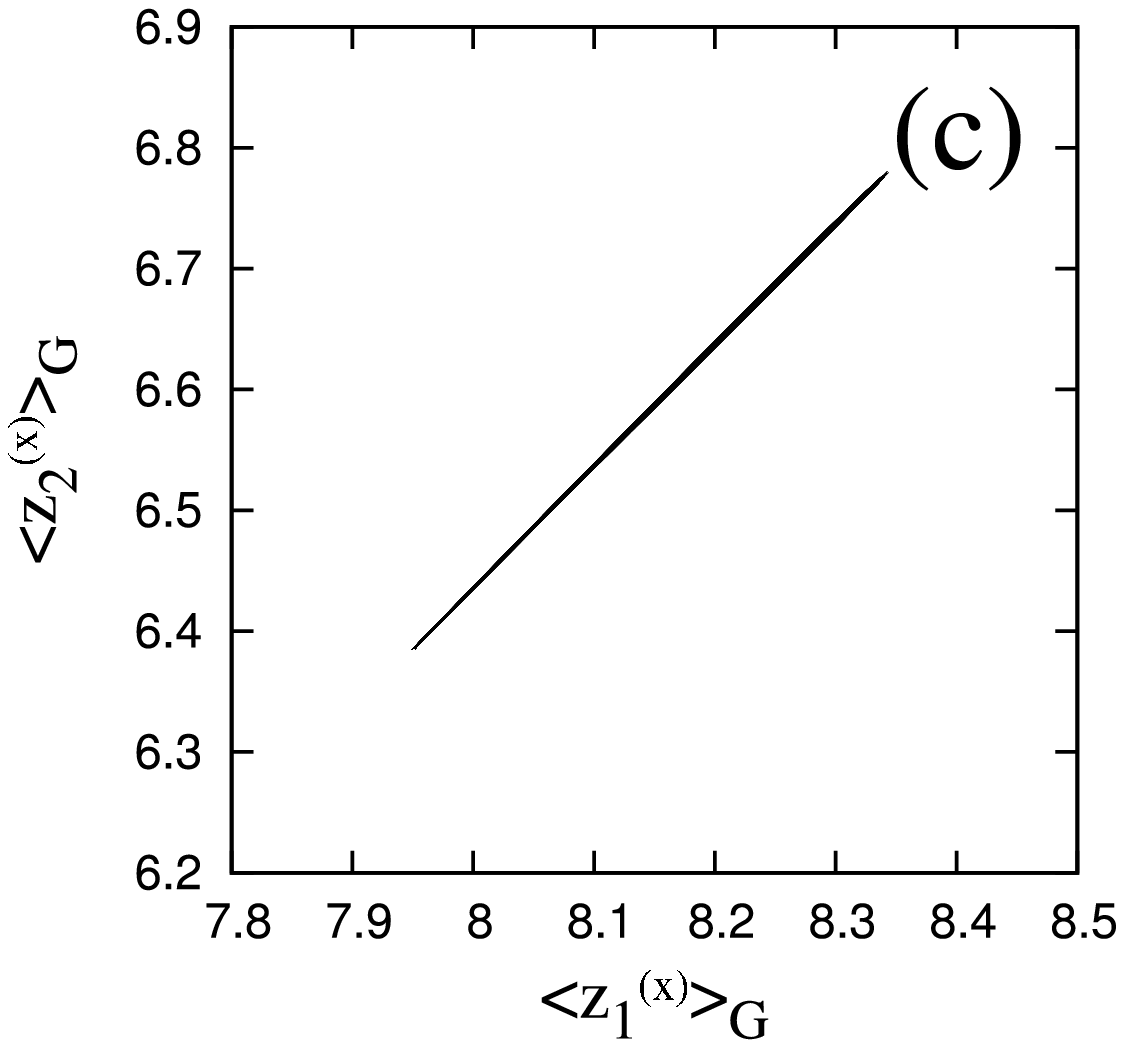}} 
    \end{tabular}
  \end{center}
 \caption{Other views in phase space. (a), (b) and (c) correspond to the attractors shown in Figs. \ref{fig:2D-ANN2_PS} (a) and \ref{fig:2D-ANN2_PS} (b), Figs. \ref{fig:2D-ANN2_PS} (c) and \ref{fig:2D-ANN2_PS} (d) and Figs. \ref{fig:2D-ANN2_PS} (e) and \ref{fig:2D-ANN2_PS} (f), respectively. It is clearly seen that the small or large mean value of $\langle z_{1}^{(x)}\rangle_{\tiny \textrm{G}}$ follows that of $\langle z_{2}^{(x)}\rangle_{\tiny \textrm{G}}$ in the synchronized state, which roughly implies so-called in-phase synchronization.} \label{fig:2D-ANN2_PS_syn}
\end{figure}

We show below the dynamics of the order parameters in phase space. The changes of types of attractors observed in terms of the LLE in Fig. \ref{fig:2D-ANN2_LLE_deltay_Dx} can be more directly seen in Figs. \ref{fig:2D-ANN2_PS} and \ref{fig:2D-ANN2_PS_syn}, where the corresponding changes of trajectories are drawn. It is seen that as the noise strength increases, the roughly estimated number of loops of chaotic trajectories decreases, while it gives rise to bifurcations from chaotic to torus type attractor. Eventually the attractor becomes that of limit cycle type, which corresponds to the occurrence of inter-cluster synchronization in the system. Within the context of rough argument, Fig. \ref{fig:2D-ANN2_PS_syn} (c) implies the occurrence of the so-called ``in-phase'' synchronization. Note that when the attractor is of torus type, the clusters are asynchronous because there are no common time period. Phenomena of noise induced chaos can be also reproduced (though not shown here), which are qualitatively in good agreement with those observed in Refs. \cite{Shiino01, Ichiki07}.

\begin{figure}[tb]
  \begin{center} 
    \begin{tabular}{l}
      \resizebox{80mm}{!}{\includegraphics[scale=1.25,angle=-90,clip]{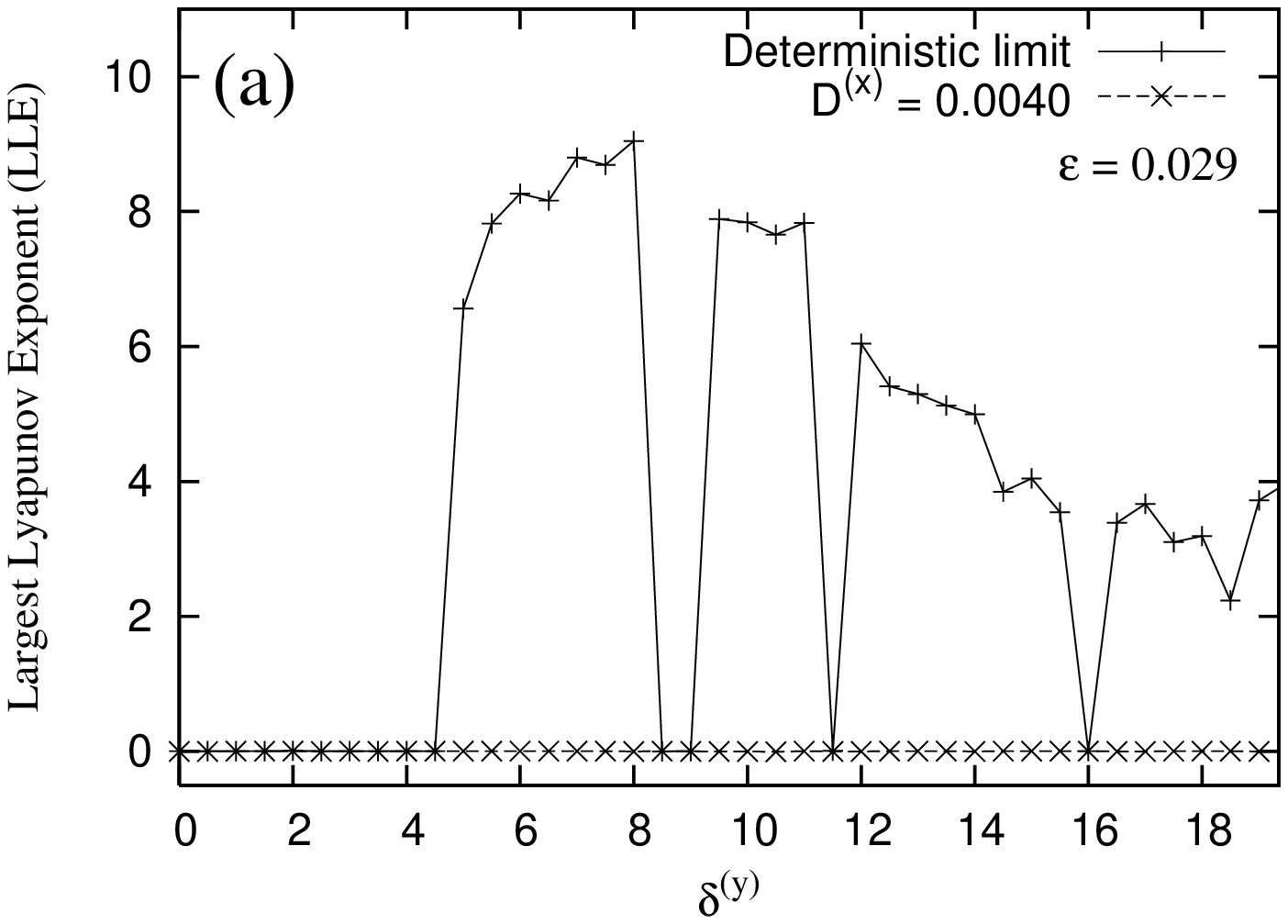}} \\
      \resizebox{80mm}{!}{\includegraphics[scale=1.25,angle=-90,clip]{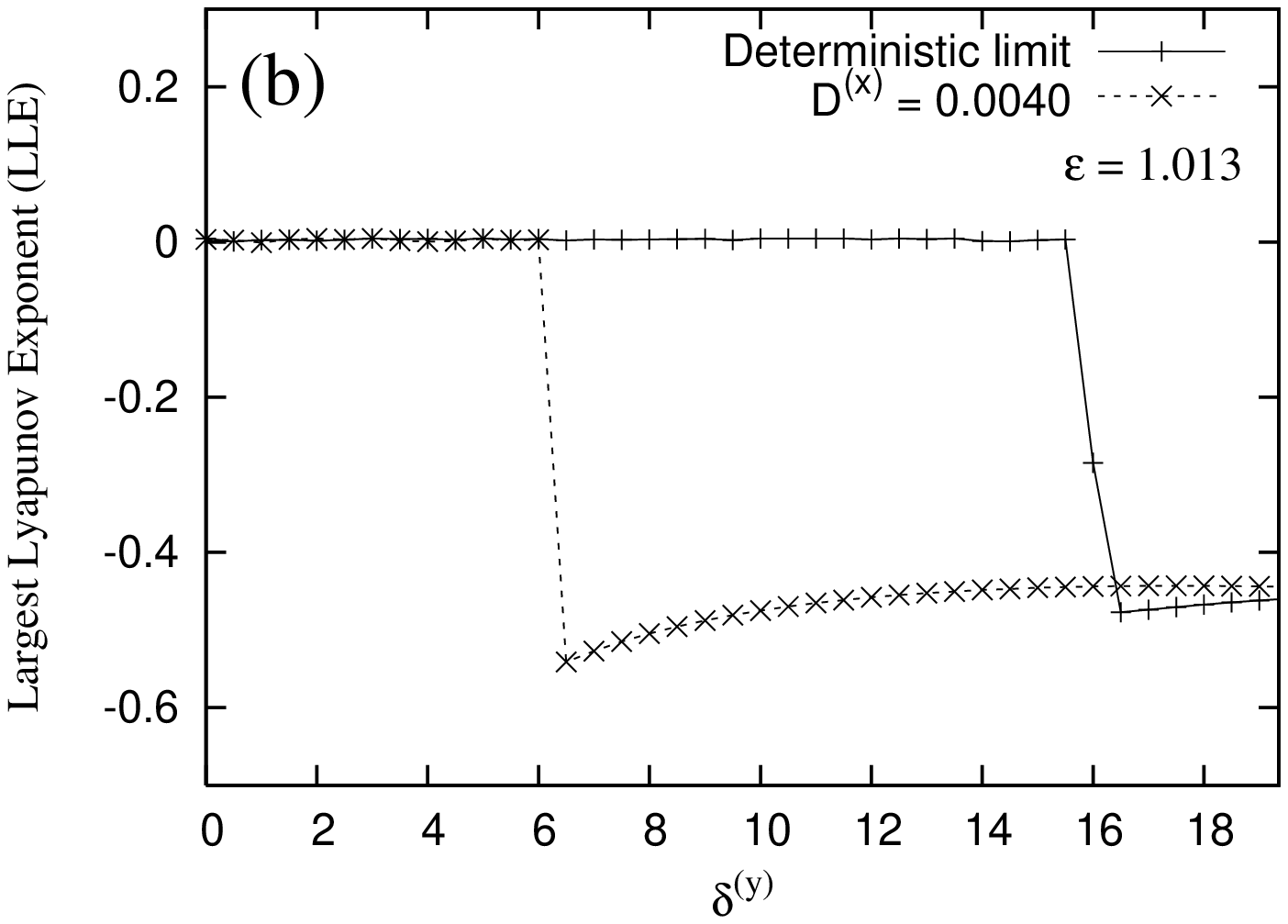}} \\
    \end{tabular}
  \end{center}
 \caption{Dependence of the largest Lyapunov exponents (LLE) on the detuning parameter $\delta^{(y)}$, when $D^{(y)}=\delta^{(x)}=0$.  (a) Weak inter-cluster coupling strength $(\epsilon=0.029)$. In the deterministic limit, the large values of $\delta^{(y)}$ induce chaotic behaviors. Under the appropriate noise intensity ({\it e.g.} $D^{(x)}=0.0040$), the chaotic attractors are suppressed to settle into limit cycle attractors. (b) Relatively strong inter-cluster coupling strength $(\epsilon=1.013)$. In the deterministic limit, the coupled oscillators do not oscillate for large values of $\delta^{(y)}$. These phenomena of oscillator death become to be observed in a broader region of $\delta^{(y)}$ for increasing noise intensity ({\it e.g.} $D^{(x)}=0.0040$). In both figures, all the points where the LLE take nearly zero show limit cycle attractors. } \label{fig:2D-ANN2_LLE_deltay_Dx_delta}
\end{figure}

We also show the effects of the frequency differences (essentially equivalent to the detuning parameter $\delta^{(y)}$) as well as the inter-cluster couping strength $\epsilon$ on synchronization phenomena with and without noise (Fig. \ref{fig:2D-ANN2_LLE_deltay_Dx_delta}). For the weak inter-cluster coupling strengths (Fig. \ref{fig:2D-ANN2_LLE_deltay_Dx_delta} (a)), it is shown that certain levels of noise intensity can give rise to synchronized states in a broad region of $\delta^{(y)}$ even if the clusters do not synchronize in the deterministic limit. In the case of the strong inter-cluster coupling strengths (Fig. \ref{fig:2D-ANN2_LLE_deltay_Dx_delta} (b)), it is seen that the quenched oscillatory states (oscillator death), which were first reported by Shiino and Frankowicz \cite{Shiino88}, are observed for larger values of $\delta^{(y)}$ in the deterministic limit. The increase of noise intensity induces breakdown of synchronized states. We note that complex behaviors can be observed in this model, depending on initial conditions (the results are not shown here). 

\section{Summary}
We have shown the effects of independent Langevin noise on the behavior of synchronization of mean-field coupled limit cycle oscillators in the thermodynamic limit. The ensemble is assumed to consist of two clusters of oscillators with different native frequencies. To investigate stochastic inter-cluster synchronization phenomena, we have used the NFPE approach. Since the NFPE for our model yields a Gaussian type probability density as its solution for large times, the time evolution equations of the relevant order parameters of the system become to be analytically derived. It should be noted that the method turns out to be a noise level-free analysis. Solving the order parameter equations numerically, we have systematically investigated the occurrence of nonequilibrium phase transitions with changes in the inter-cluster coupling strength $\epsilon$ and the detuning parameters $\delta^{(\mu)}$ responsible for native frequency difference, and the Langevin noise intensity $D^{(\mu)}$. The results have shown various interesting bifurcation phenomena including the inter-cluster synchronization and chaotic attractors induced by noise. To our knowledge, papers studying analytically systematic models of general types of limit cycles from our viewpoint and also available for comparisons with our results are very few.

Details of the relationship between the parameters involving the inter-cluster coupling strength $\epsilon$ as well as the detuning parameters $\delta^{(\mu)}$ and the behaviors of the inter-cluster synchronization will be reported elsewhere. Furthermore, the effects of the multiplicative noise including synaptic noise on the inter-cluster synchronization together with comparative discussions of the results of phase models will be also reported elsewhere.

\acknowledgments
K.O. acknowledges the financial support from the Global Center of Excellence Program by MEXT, Japan through the ``Nanoscience and Quantum Physics'' Project of the Tokyo Institute of Technology.

\end{document}